\documentstyle[12pt]{article}
\def\d{{\rm d}}
\begin{document}
\begin{titlepage}
\begin{flushright}
IISc/CTS/09/94,\\
TIFR/TH/94-41,\\
{\bf To appear in IJMP A}
\end{flushright}
\begin{center}
\vspace{0.4cm}
\Large {\bf Photon Structure Functions: Target Photon Mass Effects and 
QCD Corrections} 
\vspace{1.0cm}
\normalsize

\centerline{\large\bf Prakash Mathews$^{\dag}$} 
\centerline{\em Centre for Theoretical Studies}
\vspace{-.2cm}
\centerline{\em Indian Institute of Science}
\vspace{-.2cm}
\centerline{\em Bangalore 560 012, INDIA.}
\vspace{.2cm}
\centerline {and} 
\vspace{.2cm}
\centerline{\large\bf V. Ravindran$^{\ddag}$} 
\centerline{\em Tata Institute of Fundamental Research} 
\vspace{-.2cm}
\centerline{\em Homi Bhabha Road, Bombay 400 005, INDIA.} 
\vspace{.6cm}

\centerline{\bf Abstract}
\begin{quote}

\small{
We present a systematic analysis of the polarised and unpolarised 
processes $e^+ ~e^- \rightarrow e^+ ~e^- X$ in the deep inelastic 
limit and study the effects of target photon mass (virtuality) on 
the photon structure functions.  The effect of target photon 
virtuality manifests as new singly polarised structure functions 
and also alters the physical interpretation of the unpolarised 
structure functions.  The physical interpretation of these structure 
functions in terms of hadronic components is studied using the free 
field analysis.  We also retrieve the real photon results in the limit
the virtuality goes to zero.  Assuming factorisation of 
the photon structure tensor, the relevant QCD corrections to the 
various twist two structure functions are evaluated.  
}
\end{quote}
\end{center}

\rule{156mm}{0.3mm}
\vspace{.5cm}
\begin{flushbottom}
\begin{flushleft}
$\dag$ Address after August:\\
Tata Institute of Fundamental Research,
Homi Bhabha Road, Bombay 400 005, INDIA.

$\ddag$ Presently at:\\ 
Theory Group, Physical Research Laboratory, Navrangpura 
Ahmedabad 380 009, INDIA\\
email: ravi@prl.ernet.in
\end{flushleft} 
\end{flushbottom} 
\end{titlepage}

\section{Introduction}
The collision of photons at high energy electron-positron 
colliders is yet another comprehensive laboratory for 
testing Quantum Chromodynamics (QCD).  The process $ e^+ 
~e^- \rightarrow e^+ ~e^- X$ (hadrons) (Fig.~1), at very 
high energies can be studied in terms of the hadronic structure 
of the photon.  This process is dominated by the photon-photon 
$\rightarrow $ hadrons subprocess. 
Ahmed and Ross \cite {AR} have studied the photon structure 
function by considering the $\gamma ~\gamma$ point scattering 
and using the Operator Production Expansion (OPE) to take the 
nonperturbative effects into account.  The behaviour of this 
process in the context of perturbative QCD was first studied 
using the OPE by Witten {\cite{EW}}, who showed that the 
unpolarised structure function $F_2^\gamma$ increases as $\ln 
Q^2$, where $Q^2\equiv -q^2$ ($q$ is the momentum of the 
probing photon).  The complete next to leading order correction
to unpolarised structure functions and their phenomenological
implications can be found in \cite{EL}.  
In the recent past the polarised structure function 
has attracted a lot of attention \cite{AM,ET,SDB,NSV}.  In \cite
{AM} the OPE analysis has been extented to the polarised sector, 
while in \cite{ET,SDB} the first moment of polarised structure 
function has been evaluated and found to be zero for massless 
photons.  The sensitivity of this sum rule due to the off-shell 
nature of the target photon has been addressed in \cite{NSV}. 

Although the virtuality of the photon structure function had
been studied \cite{VPS} in the past, there has not been much emphasis on the
effect arising out of the scalar polarisation $(\lambda =0)$ of
the virtual photon ($k^2 < 0$, where $k$ is momentum of the target
photon).
{\it The off-shell 
nature of the photon gives rise to scalar polarisation which in turn
induces new structure functions to fully characterise the photon\/}. 
Some of these new structure functions are of leading twist (twist two) 
and would contribute in the Deep Inelastic Scattering (DIS) limit.
Hence a comprehensive study of the photon structure functions at this 
stage including the target mass effects and QCD corrections is much 
awaited.  

In the process we have considered, the subprocess photon-photon 
$\rightarrow$ hadrons involves a large off-shell photon probing an 
off-shell target photon in the DIS limit.  We {\it treat\/} the target 
photon as a composite object consisting of both hadronic and 
photonic components.   In our analysis, we have treated the 
photon almost on par with the nucleon.  This would imply that 
the probe photon would in fact see quarks, gluons (hadronic 
components) as well as photons (which can be produced at higher orders)
in the off-shell target photon.  All the higher order 
effects that go into the production of quarks, gluons and 
photons in the target photon are collectively treated as a blob as shown in 
the Fig.~4.  Hence we are now in a position to utilise all
the machinery that is used in the lepton-nucleon 
DIS.  Due to the off-shell nature of the target 
photon, additional polarisation of the target gives rise to 
new structure functions to completely characterise the process.  
These new structure functions are the singly polarised structure 
functions.  In the context of the photon, these new structure 
functions arising due to scalar polarisation of the target photon 
(mass effect) are being considered for the first time.  To 
understand the structure functions in terms of the hadronic 
components we perform a free field analysis of this process.  
In doing so we arrive at various sum rules and relations among 
the various photon structure functions.  These structure 
functions are related to the photon matrix elements of some
bilocal operators similar to those one comes across in DIS and 
Drell-Yan.  We assume a factorisation of hard and soft parts  
in order to calculate the hadronic and photonic contributions 
to the cross section.  Our approach is different from the 
previous analysis which usually uses either the parton model picture 
or OPE.

The rest of the paper is organised as follows.  In section 2 
we study the effects of the virtuality of the target photon 
on the photon structure tensor and hence the polarised and 
unpolarised cross sections.  In section 3 we perform a free 
field analysis to study the physical interpretation of the 
new structure functions and demonstrate how the unpolarised
structure functions are altered due to the virtuality of the
photon.  In section 4 we use the factorisation method to
evaluate the QCD corrections to the new photon structure
functions.  Finally we conclude in section 5.

\section{Target photon mass effects}
Consider the process $e^- (p_1,s_1)~ e^+ (p_2,s_2) \rightarrow e^- 
(p_1^\prime) ~e^+ (p_2^\prime)~ X(p_{\bf n})$, where $s_1$, $s_2$ are 
the polarisation vectors of the leptons  and $X$ represents the final 
state hadrons.  In the $cm$ frame, the momenta of the in coming and 
out going particles are parametrised as $p_1=(E_1,0,0,E_1)$, 
$p_2=(E_1,0,0,-E_1)$ and $p_1^\prime = E_1^\prime (1,0,\sin \theta_1,
\cos \theta_1)$, $p_2^\prime =E_2^\prime(1,0,-\sin \theta_2,-\cos
\theta_2)$ respectively.  $\theta_{1,2}$ are the scattering angles of out 
going leptons with respect to the beam axis. This process (Fig. 1) at
very high energies is 
dominated by the photon-photon $\rightarrow$ hadrons subprocess.  In 
our analysis we consider a probe photon $\gamma^* (q)$ probing a target
photon $\Gamma(k)$ in the DIS limit.  We keep the target photon in 
general to be nonperturbative and study the effect of virtuality
on the photon structure functions.  The total cross section for this 
process is given by
\begin{eqnarray}
\d \sigma=\frac{1}{8 E_1^2}~\frac{\d^3p_1^\prime}{(2\pi)^3 2 E_1^\prime}
~\frac{\d^3p_2^\prime}{(2\pi)^3 2 E_2^\prime}~\prod_{i=1}^n
~\frac{\d^3p_i}{(2\pi)^3 2 E_i} ~\vert T_n \vert^2 ~(2 \pi)^4 \delta^4(
p_1+p_2-p_1^\prime-p_2^\prime-p_n)~,
\end{eqnarray}
where $T_n$ is the transition amplitude and $n$ refers to the final
state.  The double differential cross section is found to be  
\begin{eqnarray}
\frac{\d \sigma^{s_1 s_2}}{\d x~\d Q^2} &=& \frac{\alpha^3}
{4 x s^2 Q^2} 
\int \frac{{\d \kappa^2}}{\kappa^2}
 ~ \int_x^1 \frac{\d y}{y^2}~ L^{\mu \mu
^\prime} (q,p_1,s_1)~ L^{\nu \nu^\prime} (k,p_2,s_2) 
\nonumber \\
&& \times \sum_{\lambda=0,\pm 1} g^{\lambda \lambda} ~\epsilon^*_
\nu (k,\lambda) ~\epsilon_{\nu^\prime} (k,\lambda)~ W^\Gamma_{\mu 
\mu^\prime} (q,k,\lambda) ~,
\label{CS}
\end{eqnarray} 
where $\alpha = e^2/4\pi$, $s$ is the centre of mass energy of 
incoming leptons, $q=p_1-p_1^\prime$ and $k=p_2-p_2^\prime$ 
are the momenta of the probe and target photons with invariant 
mass $Q^2=-q^2$ and $\kappa^2=-k^2$ respectively.  
The Bj\"orken variable with 
respect to the target positron is defined as $x \equiv Q^2/2 
\tilde \nu$, where $\tilde \nu= p_2 \cdot q$ and that with respect to 
the target photon is $y \equiv Q^2/2 \nu$, where $\nu= k \cdot q$. 
The DIS limit corresponds to $Q^2$, $\tilde \nu$, $\nu \rightarrow
\infty$ with $x$ and $y$ fixed.   
Further we are interested in the region $\kappa^2 \ll Q^2$ and hence do not consider
terms of the order ${\cal O} (\kappa^2/Q^2)$.  However we would like to address the
effects of scalar polarisation on the photon structure functions.
The vector $\epsilon^\mu(k,\lambda)$ is the polarisation vector of the target
photon with polarisation $\lambda$.  
Photons with $k^2 < 0$, are
characterised by $\lambda=0$ (scalar) in addition to $\lambda=\pm 1$
(transverse) polarisation states.  The polarisation vectors
corresponding to these states satisfy the relation:
$\sum_{\lambda=0,\pm 1} g^{\lambda \lambda} \epsilon_\mu^*(k,\lambda)
\epsilon_\nu(k,\lambda)=k^2 g_{\mu\nu}-k_{\mu} k_{\nu}$ with 
$\epsilon(k,\lambda) \cdot k=0$.
The lepton tensor $L^{\mu \nu}(q,p,s_i)$ is generically defined as
\begin{eqnarray}
L^{\mu \nu}(q,p,s_i) = 4~ p^\mu p^\nu - 2~(p^\mu q^\nu + p^\nu q^\mu)
+ 2~ p \cdot q~ g^{\mu \nu} - 2i~ \epsilon^{\mu \nu \alpha \beta}~ s^i_\alpha
~q_\beta ~,
\label{LT}
\end{eqnarray}
where $s_i \cdot p=0$ and $s_i^2=-m_e^2$.
The cross section (eqn.~(\ref{CS})) can be written in terms 
of the electron structure tensor $W^e_{\mu \mu^\prime}(q,p_2,s_2)$ as
\begin{eqnarray}
\frac{\d \sigma^{s_1 s_2}}{\d x ~ \d Q^2} =\frac{\alpha^2}{4 x^2 s^2 Q^2}~ 
L^{\mu \mu^\prime} (q,p_1,s_1)~ W^e_{\mu \mu^ \prime} (q,p_2,s_2) ~,
\label{TCS}
\end{eqnarray}
where $W^e_{\mu \mu^\prime}(x,Q^2,s_2)$ is
\begin{eqnarray}
W^e_{\mu \mu^\prime}(x,Q^2,s_2)=\alpha x 
\int \frac{{\d \kappa^2}}{\kappa^2}~ \int^1_x \frac{\d y}{y^2} 
L^{\nu \nu^\prime} (k,p_2,s_2) \sum_{\lambda=0,
\pm 1} g^{\lambda \lambda}~ \epsilon^*_{\nu} (k,\lambda)~\epsilon
_{\nu^\prime} (k,\lambda)~ W^\Gamma_ {\mu \mu^\prime} (q,k,\lambda)~.
\label{EST}
\end{eqnarray}
Due to the presence of the nonperturbative photon tensor $W_{\mu \nu}^
\Gamma(q,k,\lambda)$, the electron structure tensor $W_{\mu \nu}^e(x,
Q^2,s_2)$ can at best be parametrised as a spin $1/2$ target in terms of 
vectors $q, p_2, s_2,$ subject to the symmetries as
\begin{eqnarray}
W^e_{\mu \nu}(x,Q^2,s_2) & =& F_1^e(x,Q^2)~ G_{\mu \nu}  + F_2^e(x,Q^2) 
~\frac{\widetilde R_\mu\widetilde  R_\nu}{\tilde \nu}
\nonumber\\
&& + \frac{i}{\tilde \nu^2} \epsilon_{\mu \nu \alpha \beta} 
\left ( g_1^e(x,Q^2)~ \tilde \nu ~q^\alpha s_2^\beta + g_2^e(x,Q^2)~  
q^\alpha \left(\tilde \nu~ s_2^\beta - s_2 \cdot q~ p_2^\beta\right)\right ) ~.
\label{EST2}
\end{eqnarray}
Here $F^e_{1,2} (x,Q^2)$ and $g^e_{1,2} (x,Q^2)$ are the unpolarised
and polarised electron structure functions respectively.  The tensor 
coefficients in the above equation are defined as
\begin{eqnarray}
G_{\mu \nu} = -g_{\mu \nu} +\frac{q_\mu q_\nu}{q^2}~ , \qquad 
 \widetilde R^\mu = p_2^\mu -\frac{\tilde \nu}{q^2} ~q^\mu ~.
\end{eqnarray} 

The photon structure tensor $W^\Gamma_{\mu \nu} (q,k,\lambda)$ is the
imaginary part of the forward amplitude $\gamma^*(q,\lambda^\prime)$ 
$\Gamma(k,\lambda) \rightarrow \gamma^*(q,\lambda^\prime)$ $\Gamma(k,
\lambda)$ and can be defined as the Fourier transform of the commutator 
of electromagnetic ($em$) currents $J_\mu(\xi)$ sandwiched between 
target photon states, as
\begin{eqnarray}
W_{\mu \nu}^\Gamma(k,q,\lambda) &=& \frac{1}{2 \pi} \int \d^4 \xi ~ 
e^{- i q\cdot \xi} ~\langle \Gamma(k, \epsilon^*(\lambda)) \vert ~ 
\left [J_{\mu}(\xi), J_{\nu}(0) \right ]~ \vert \Gamma(k,\epsilon
(\lambda)) \rangle_c ~,
\label{FTD}
\end{eqnarray}
where the subscript $c$ denotes the connected part.  For a real photon $
\Gamma(\lambda) \rightarrow \gamma(\lambda)$, the polarisation states are
$\lambda = \pm1$ while the probe photon polarisation $\lambda^\prime =0, 
\pm 1$, since $q^2 \ne 0$ and is very large in the DIS limit.  The number of 
independent helicity amplitudes is four, corresponding to the four structure 
functions $F_{1,2}^\gamma (y,Q^2)$ and $g_{1,2}^\gamma (y,Q^2)$.  The virtual 
photon $(\kappa^2 \ne 0)$ on the other hand is 
characterised by polarisation states $\lambda = 0,\pm 1$.  Enumerating the
number of independent helicity amplitudes subject to parity and time
reversal invariance gives eight independent amplitudes.  Hence the photon
structure tensor $W^\Gamma_{\mu \nu} (k,q,\lambda)$ can be parametrised 
in a gauge invariant way in terms of the eight structure functions using 
general symmetry arguments such as time reversal invariance, parity, 
hermiticity and current conservation. Thus
\begin{eqnarray}
W^\Gamma_{\mu \nu} (y,Q^2,\kappa^2,\lambda)\!\! &=&\!\! \frac{1}{\kappa^4}~ 
\left \{ F^\Gamma_1 (y,Q^2,\kappa^2) ~G_{\mu \nu} + F^\Gamma_2 (y,Q^2,
\kappa^2) ~\frac{ R_\mu  R_\nu}{\nu} \right .
\nonumber \\
&& +~ b^\Gamma_1(y,Q^2,\kappa^2) ~ r_{\mu \nu} 
+ ~b^\Gamma_2(y,Q^2,\kappa^2) ~ s_{\mu \nu} 
\label{PST} \\
&& +~ b^\Gamma_3(y,Q^2, \kappa^2) ~ t_{\mu \nu}  +~ b^\Gamma_4(y,Q^2,
\kappa^2) ~ u_{\mu \nu} 
\nonumber\\
&&\left . +~ \frac{i}{\nu^2}~ \epsilon_{\mu \nu \lambda \rho}~ 
\left ( g^\Gamma_1(y,Q^2,\kappa^2) \nu ~q^\lambda~  s^\rho 
+~ g^\Gamma_2(y,Q^2,\kappa^2) ~q^\lambda~ ( \nu ~ s^\rho - s 
\cdot q ~k^\rho) \right ) \right \}~,
\nonumber
\end{eqnarray}
where $s_\mu$ is the spin vector of the target photon, $R_\mu=x
~\widetilde R_\mu/y$ and the various tensors are defined as 
\begin{eqnarray}
 r_{\mu \nu} &=&\frac{1}{4}  \left (\frac{k\cdot E^* ~k\cdot E}{\kappa^4}
-  ~\bar \alpha^2 \right )~ G_{\mu \nu} ~, \qquad 
 s_{\mu \nu} = \frac{1}{4 \nu} ~\left (\frac{k\cdot E^* ~k\cdot E}{\kappa^4} 
-  ~\bar \alpha^2 \right ) ~ R_\mu ~ R_\nu  ~,\nonumber\\
 t_{\mu \nu} &=&\!\!\! -\frac{1}{8 \nu} \left (\frac{k\cdot E^*} {\kappa^2} ~(R_\mu ~E_\nu
+ R_\nu ~E_\mu) + \frac{k\cdot E}{\kappa^2} ~(R_\mu ~E_\nu^*
+ R_\nu ~E_\mu^*) - 4~(1-\bar \alpha^2) R_\mu R_\nu \right ),
\nonumber\\
u_{\mu \nu} &=& \frac{1}{4 \nu} \left (E^*_\mu ~E_\nu + E^*_\nu ~E_\mu 
+ 2~\kappa^2 ~G_{\mu \nu} + 2~(1-\bar \alpha^2) R_\mu ~R_\nu 
\right ) ~, \\
E_\mu &=& \epsilon_\mu - \frac{q \cdot \epsilon}{\nu} ~k_\mu ~, 
\quad \qquad \bar \alpha^2 = 1- \frac{\kappa^2 Q^2}{\nu^2} ~, \quad \qquad 
s^\mu \equiv \frac{i}{\kappa^2}~\epsilon^{\mu \nu \alpha \beta} ~\epsilon^*_
\nu ~\epsilon_\alpha ~k_\beta~. 
\nonumber
\end{eqnarray}
This is the standard decomposition of a massive spin one composite
target \cite{HJM}.  The additional input that has gone in the case
of the virtual photon is the manifest gauge invariance of the tensor
coefficients.
The four new structure functions $b^\Gamma_{1-4} ~(y,Q^2,\kappa^2)$ 
are due to the off-shell nature ($\lambda =0$, scalar polarisation ($k^2 <0$)) 
of the target photon.  The tensor coefficients of these additional structure
functions vanish when the target photon polarisation is summed, and survive when
the target is polarised while the probe polarisation is summed.  Hence
these structure functions are called the singly polarised structure functions.  This singly 
polarised nature is characteristic of a spin one target \cite{HJM}.
In the context of photon structure functions which are realised in a 
$e^+ ~e^- \rightarrow e^+ ~e^- X$ process, the singly polarised part
does not manifest itself as in a spin one target, but turns out to be
a part of the unpolarised cross section.  

The unpolarised, singly polarised and the polarised sectors which 
characterise the virtual photon tensor $W^\Gamma_{\mu \nu} (y,Q^2,
\kappa^2)$ in eqn.~(\ref{PST}) are all independent and can be extracted using the 
following combinations of polarisation states
\begin{eqnarray}
\overline W_{\mu \nu} &=& \sum_{\lambda=0,\pm 1} g^{\lambda \lambda} 
~W^\Gamma_{\mu \nu} (\lambda), 
\nonumber\\
\delta W_{\mu \nu} &=& \sum_{\lambda=0,\pm 1} C(\lambda) ~W^\Gamma_{\mu \nu} 
(\lambda),
\label{PC}\\
\Delta W_{\mu \nu} &=& \sum_{\lambda=0,\pm 1} C^\prime (\lambda)~W^\Gamma_
{\mu \nu} (\lambda),
\nonumber
\end{eqnarray}
where $g^{\lambda \lambda}$ is the metric tensor, $C(\lambda) =2$ for $\lambda=0$; 
$-1$ for $\lambda=\pm 1$ and $C^\prime (\lambda) =0$ for $\lambda=0$; $\pm 1$ 
for $\lambda= \pm 1$. 
The polarised combination $\Delta W_{\mu \nu}$ is same as the real photon combination.
This is due to the fact that the virtuality of the photon induces the $\epsilon_
\mu (\lambda=0)$ polarisation state and this does not contribute to the
antisymmetric polarised part $\Delta W_{\mu \nu}$.  In contrast the 
unpolarised combination $\overline W_{\mu \nu}$ is altered due to the 
virtuality as it is a sum over all the polarisation states of the virtual 
photon.  For the real photon the $\epsilon_\mu (\lambda=0)$ polarisation
state would be absent and hence the unpolarised $\overline W_{\mu \nu}$
and singly polarised $\delta W_{\mu \nu}$ combination reduces to the real
photon unpolarised combination which is the sum of $\lambda= \pm 1$ 
polarisation states.

The unpolarised and polarised cross sections for $e^+ e^- \rightarrow 
e^+ e^- X$ can be derived using eqns.~(\ref{TCS},\ref{EST}) and 
eqn.~(\ref{PST}).  The details of this derivation are given
in the Appendix.  The unpolarised cross section is given by
\begin{eqnarray}
\frac{\d \sigma^{\uparrow \uparrow +\uparrow \downarrow}}{\d x~\d Q^2}
&=& \frac{\alpha^3}{x s^2 Q^2} L^{\mu \mu^\prime}_{sym} (q,p_1,\uparrow)
\int \frac{{\d \kappa^2}}{\kappa^2} \int_x^1 \frac{\d y}{y^2} \kappa^4 
\left\{- \frac{y}{x} {\overline P}_{\gamma e}\left (\frac {x}{y} 
\right ) 
{\overline W}^\Gamma_{\mu \mu^\prime} (y,Q^2,\kappa^2) \right . 
\nonumber \\
&&\left . + ~\frac{y}{2 x} ~\delta  P_{\gamma e} \left (\frac {x}{y}
\right ) 
~\delta W^\Gamma_{\mu \mu^\prime} (y,Q^2,\kappa^2) \right\} ~,
\label{UPCS} 
\end{eqnarray}
where $\uparrow (\downarrow)$ denotes the polarisation of an electron
along (opposite) to the beam direction and $L^{\mu \mu^\prime}_{sym}
(q,p_1,\uparrow)$ is the symmetric part of the lepton tensor eqn.~(\ref{LT}).
In the above equation, the first term in the curly bracket corresponds 
to unpolarised structure function and the second term to the singly 
polarised structure functions.  The modified splitting functions 
($k^2 \not= 0$) are given by
\begin{eqnarray}
\overline P_{\gamma e}\left (\frac{x}{y} \right ) =\frac{y}{x}\left
(2-2 \frac{x}{y} + \frac{x^2}{y^2}\right)-2 ~\frac{y}{x}\left(1-\frac
{x}{y}\right) ~, \\
\delta P_{\gamma e} \left (\frac{x}{y} \right ) =\frac{y}{x}
\left(2-2 \frac{x}{y}+ \frac{x^2}{y^2}\right)-4 ~\frac{y}{x}\left
(1-\frac{x}{y}\right)  ~.
\end{eqnarray}
The first term in the above equations is the usual Weizs\"acker-Williams 
splitting function 
arising from the splitting of $e^+$ into transverse photons ($k^2 
= 0$).  The additional term arises from the emission of scalar 
polarised photon. The unpolarised splitting function $\overline P_
{\gamma e} \propto \sum_{\lambda = 0 \pm 1} g^{\lambda \lambda} 
\epsilon^*_\mu (\lambda)$ $ \epsilon_\nu (\lambda) L^{\mu \nu} 
(k,p_2,s_2)$ and the singly polarised $\delta P_ {\gamma e} ~\propto ~\sum_
{\lambda = 0 \pm 1} ~C(\lambda) ~\epsilon^*_\mu (\lambda) \epsilon_\nu 
(\lambda)$ $L^{\mu \nu} (k,p_2,s_2)$.  Hence there is a neat factorisation
of the cross section (eqn.~(\ref{UPCS})) in terms of the combination of 
polarisation states used to extract unpolarised and singly polarised
structure functions.  The polarised cross section is given by
\begin{eqnarray}
\frac{\d \sigma^{\uparrow \uparrow -\uparrow \downarrow}}{\d x~\d Q^2} 
&=& -\frac {\alpha^3}{x~s^2~Q^2}~ L^{\mu \mu^\prime}_{asym} (q,p_1,\uparrow)
\int \frac{{\d \kappa^2}}{\kappa^2}~ \int_x^1\frac{\d y}{y^2}~ \kappa^4 ~
\frac{y}{x}~ \Delta P_{\gamma e}\left(\frac{x}{y} \right) 
~\Delta W^\Gamma_{\mu \mu^\prime} (y,Q^2,\kappa^2) ~,
\label{PCS} 
\end{eqnarray}
where $L^{\mu \mu^\prime}_{asym} (q,p_1,\uparrow)$ is the antisymmetric 
part of eqn.~(\ref{LT})
and the polarised splitting function is 
\begin{eqnarray}
\Delta P_{\gamma e}\left(\frac{x}{y}\right) =2 - \frac{x}{y} ~.
\nonumber
\end{eqnarray}
Note that the splitting function $\Delta P_{\gamma e}$ is the same as 
in the real case.  As expected the additional scalar polarisation of the 
photon does not contribute to the polarised cross section.

The electron structure functions can now be related to the photon 
structure functions by substituting the photon structure tensor 
eqn.~(\ref{PST}) in eqns.~(\ref{UPCS},\ref{PCS}) and comparing it with the 
total cross section eqn.~(\ref {TCS}) after substituting eqn.~(\ref
{EST2}).  Hence we get 
\begin{eqnarray}
F_1^e(x,Q^2) &=& 2 \alpha 
\int \frac{\d \kappa^2} {\kappa^2}~ \int_x^1 \frac{\d y}{y}~
\left[ \overline P_{\gamma e}\left(\frac{x}{y}\right)~ 
F_1^\Gamma(y,Q^2, \kappa^2) \right. 
\nonumber\\ 
&&\left.- ~ \frac{1}{2} \delta P_{\gamma e} \left(\frac{x}{y} \right)~ \left(
\frac{\bar \alpha^2}{2}~ b_1^\Gamma(y,Q^2,\kappa^2) - \frac{\kappa^2}{2 \nu}~ 
b_4^\Gamma (y,Q^2,\kappa^2)\right)
\right]~, \label{F1}\\
F_2^e(x,Q^2) &=& 2 \alpha 
\int \frac{\d \kappa^2} {\kappa^2}~ \int_x^1 \frac{\d y}{y}~ 
\frac{x}{y} \left[ \overline P_{\gamma e}\left(\frac{x}{y} 
\right)~ F_2^\Gamma(y,Q^2,\kappa^2) 
\right.\label{F2} 
\nonumber\\ 
&& - ~\frac{1}{2} \delta P_{\gamma e} \left(\frac
{x}{y} \right) \left(\frac{\bar \alpha^2}{2}~ b_2^\Gamma(y,Q^2,\kappa^2) 
-~(1- \bar \alpha^2)~b_3^\Gamma (y,Q^2,\kappa^2)~\right.
\nonumber \\
&&\left. \left. +\frac{1-\bar \alpha^2}{2 \bar \alpha^2} ( 1-2\bar \alpha^2)~ 
b_4^\Gamma(y,Q^2,\kappa^2)\right)\right]~, \\
g_1^e (x,Q^2) &=& 4 \alpha \int \frac{\d \kappa^2} {\kappa^2}~ 
\int_x^1 \frac{\d y}{y}~ 
\left[ \Delta P_{\gamma e}\left( \frac{x}{y}\right)~ g_1^\Gamma
(y,Q^2, \kappa^2) \right ]~.\label{G1} 
\end{eqnarray}
Note that the unpolarised electron structure functions are related to the 
singly polarised structure functions of photon.  This extra contribution comes 
from the scalar polarisation (virtuality) of the photon which is also 
reflected in the modified splitting functions.  To leading order in the 
DIS limit, the unpolarised electron structure 
functions $F_{1,2}^e (x,Q^2)$ are modified only by the twist two singly 
polarised structure functions $b_{1,2}^\Gamma (y,Q^2,\kappa^2)$ respectively.  
The other singly polarised structure functions $b_{3,4}^\Gamma(y,Q^2,\kappa^2)$
do not contribute in the DIS limit as their coefficients are of the
form $(1- \bar \alpha^2)$ which in this limit goes like
$\kappa^2/Q^2$. This naively counts the twist of the new
structure functions i.e., $b^\Gamma_{1,2}(y,Q^2,\kappa^2)$ are twist two
and $b^\Gamma_{3,4}(y,Q^2,\kappa^2)$ are twist four structure functions.
Higher twist contributions go as powers of $\kappa^2/Q^2$ and hence
the structure functions $b^\Gamma_{3,4}(y,Q^2,\kappa^2)$ are
not pursued any further.
It is imperative at this stage to show that in the limit $\kappa^2
\rightarrow 0$, we can retrieve the real photon results.  At this stage
we are not in a position to restore the real photon results as we do
not know the behaviour of the structure function in the limit $\kappa^2
\rightarrow 0$.  So in the next section we
show that our present analysis is consistent with the earlier works
on photon structure function in the $\kappa^2 \rightarrow 0$ limit (real
photon).  We also show that two of the new structure functions
($b^\Gamma_{1,2}(y,Q^2,\kappa^2)$) are of twist two. 
The polarised structure function $g_1^e (x,Q^2)$ on the other hand is
unaffected by the virtuality of the photon.

In terms of the above relations we can compute the unpolarised and 
polarised cross sections and these are given by
\begin{eqnarray}
\frac{\d \sigma^{\uparrow \uparrow +\uparrow \downarrow}}{\d x~\d Q^2}&=& 
\frac{\alpha^2}{x^2~ s~ Q^2} \left\{F_1^e(x,Q^2) ~\frac {Q^2}{s} - F^e
_2(x,Q^2) \left (1 - \frac{xs}{Q^2} \right ) \right\}~,\\
\frac{\d \sigma^{\uparrow \uparrow -\uparrow \downarrow}}{\d x~\d Q^2}&=& 
2\frac {\alpha^2}{x~ s~ Q^2} ~g_1^e(x,Q^2) \left(1 -\frac{Q^2}{2xs} 
\right)~.
\end{eqnarray}
Now by substituting for the electron structure functions from 
eqns.~(\ref{F1}--\ref{G1}) the $n^{\rm th}$ moment of the above differential
cross section can be related to the $(n-1)^{\rm th}$ moment of the 
photon structure functions.  

In this section we have shown that in the DIS limit, the virtuality
of the  target photon manifests as new singly polarised twist two 
structure functions $b_{1,2}^\Gamma (x,Q^2,\kappa^2)$ and in addition
the $e \rightarrow \gamma$ splitting function also gets modified.
Our next task is to understand these new structure functions in terms 
of the parton distributions.

\section{Free field analysis}
To understand the hadronic structure of the photon structure functions,
we make use of the free field analysis {\it akin\/} to the one used in the 
case of nucleon structure function.  We present a systematic study of 
these new structure functions i.e., their twist structure and their 
physical interpretation in terms of parton content.  The twist analysis 
is necessary as the partonic interpretation is possible only for twist 
two operators.  This is done using free field analysis.  
We will show from the twist analysis that $b^\Gamma_{1,2}(y,Q^2,\kappa^2)$ have
definite parton model interpretation as they are related to twist two
operators.  We will also show as a by product that in the $\kappa^2
\rightarrow 0$ limit, we can reproduce the real photon results. 

In this analysis one assumes the $em$ current $J_\mu$ to be made of free 
quark currents while the photon states are nonperturbative.  This analysis
should therefore not be confused with the conventional Quark Parton Model in 
the context of photon structure function wherein the photon is directly
coupled to the charge of the bare quark and is realised only at large $\kappa^2$.
In the DIS limit, the leading contribution
to the commutator in eqn.~(\ref {FTD}) comes from the light
cone region $\xi^2 \rightarrow 0$.  Noting that the commutator is proportional
to the imaginary part of the time ordered product of currents and using 
Wick's expansion, we get
\begin{eqnarray}
[J_\mu (\xi),J_\nu(0)] &=& \frac{\delta^{(1)}~ (\xi^2)~ \xi^\lambda}{\pi}
\left \{ \sigma_{\mu \lambda \nu \rho}~ O^\rho_{(-)} (\xi) - i \epsilon_
{\mu \lambda \nu \rho}~ O^\rho_{(+)5} (\xi) \right \} ~,
\label{CC}
\end{eqnarray}
where
\begin{eqnarray}
\delta^{(1)}(\xi^2) & = & \frac{\partial}{\partial \xi^2}
\delta(\xi^2) ~, \nonumber\\
\sigma_{\mu \lambda \nu \rho} &=& g_{\mu \lambda}~ g_{\nu \rho} - 
g_{\mu \nu}~ g_{\lambda \rho} + g_{\mu \rho}~ g_{\lambda \nu} ~,\nonumber\\  
O_{(\pm)}^\rho (\xi) &=& :\overline \psi (\xi)~ \gamma^\rho~ \psi (0)~ \pm~ 
\overline \psi (0)~ \gamma^\rho~ \psi (\xi): ~,\\
O_{(\pm)5}^{\rho}(\xi) &=& :\overline \psi (\xi)~ \gamma^\rho~ \gamma_5~
\psi (0)~ \pm~ \overline \psi (0) ~\gamma^\rho ~\gamma_5~ \psi (\xi):
~,\nonumber
\end{eqnarray}
where the symbol : : implies normal ordering of operators.  Substituting 
this in eqn.~(\ref{FTD}), performing the 
$\d^4 \xi$ integral and comparing the tensor coefficients with 
eqn.~(\ref{PST}), we relate the structure functions to various scaling 
functions as given in Table 1.  $\widetilde A (y)$, $\widetilde B 
(y)$ and $\widetilde C (y)$ in the Table 1 are defined as 
\begin{eqnarray}
{\displaystyle \int} \d z^\prime ~ e^{iz^\prime k\cdot\xi} \pmatrix {\widetilde A (z^\prime ) \cr \widetilde B (z^\prime ) 
\cr \widetilde C (z^\prime ) } = \pmatrix {A (k\cdot\xi) \cr B (k\cdot\xi) \cr C (k\cdot\xi) } =
\sum_n \frac{1}{(n+1)!}~  k\cdot\xi^{n-1} \pmatrix{ (n+1) A_n ~ k\cdot\xi \cr B_n 
\cr C_n ~ k\cdot\xi} ~,
\end{eqnarray}
where $A_n$~, $B_n$~, $C_n$ are the expansion coefficients 
of local photon matrix elements given below
\begin{eqnarray}
\langle \Gamma (k, \epsilon^*) \vert O_{(-)}^{\rho \mu_1 \cdots \mu_n} (0) 
\vert \Gamma (k, \epsilon) \rangle \!\! &=& \!\! 2 A_n {\cal S} (k^\rho k^{\mu_1} 
\cdots k^{\mu_n}) + B_n {\cal S} \left [ \left ({\epsilon^\rho}^* 
\epsilon^{\mu_1} + {k^\rho k^{\mu_1}} \right ) k^{\mu_2} \cdots 
k^{\mu_n} \right ],\\
\langle \Gamma(k, \epsilon^*) \vert O_{(+)5}^{\rho \mu_1 \cdots \mu_n} (0) 
\vert \Gamma(k, \epsilon) \rangle &=& C_n {\cal S} (s^\rho k^{\mu_1} \cdots 
k^{\mu_n}) ~.
\end{eqnarray}
Here $A_n, B_n, C_n$ are functions of Lorentz invariants such as 
$k^2, s^2$ etc.  ${\cal S}$ denotes symmetrisation with respect 
to all indices.  This is done to ensure that only leading twist operators 
contribute.  The matrix element $A_n$ contributes to the unpolarised 
part, $B_n$ to the singly polarised part and $C_n$ to the polarised part.  
{}From Table 1 it is clear that both the unpolarised structure 
functions $(F^\Gamma_{1,2}~ (y))$ and the singly polarised structure 
functions $(b^\Gamma_{1,2}~ (y))$ satisfy Callan-Gross relation.  
In addition we find that $b^\Gamma_3(y)$ and $b^\Gamma_4(y)$ are related 
to $b^\Gamma_2 (y)$ by the following relations:
\begin{eqnarray}
b^\Gamma_4(y) &=& - \int_y^1 \frac{\d y^\prime}{y^\prime} ~b^\Gamma_3
(y^\prime)~,\\
b^\Gamma_3(y) &=& - \int_y^1 \frac{\d y^\prime}{y^\prime} ~b^\Gamma_2
(y^\prime)~.
\end{eqnarray}

The physical interpretation of these structure functions can be given
using the above free field analysis.  This is done by substituting the 
current commutator eqn.~(\ref{CC}) in eqn.~(\ref{FTD}) and performing 
only the $ \d \xi^+$ and $\d \xi_\perp$ integrals, where $\xi^\pm = (\xi
^0 \pm \xi^3)/\sqrt 2$, $\xi_\perp = (\xi^1,\xi^2)$.  The unpolarised, 
singly polarised and polarised structure functions can be separated 
using the combinations of the polarisation states defined in eqn.~(\ref{PC}).
Using appropriate projection operators for the various 
structure functions, we get
\begin{eqnarray}
F^\Gamma_1(y) &=&  ~\frac{1}{4\pi} \int \d \xi^- e^{-iy k^+ \xi^-} 
\langle \Gamma (k) \vert ~\overline O^{~+}_{(-)} (0,\xi^-,0_\perp) 
\vert \Gamma (k) \rangle~,\\
b^\Gamma_1(y) &=& ~\frac{1}{2\pi} \int \d \xi^- e^{-iy k^+ \xi^-} 
\langle \Gamma(k , \epsilon^*)\vert~ \delta O^+_{(-)} (0,\xi^-,0_\perp) \vert 
\Gamma(k,\epsilon) \rangle~,\\
g^\Gamma_1(y) &=& \frac{1}{4\pi} \int \d \xi^- e^{-iy k^+ \xi^-} 
\langle \Gamma(k,\epsilon^*)\vert~ \Delta O^+_{(+)5} (0,\xi^-,0_\perp) 
\vert \Gamma(k,\epsilon) \rangle ~,
\end{eqnarray}
where the superscript $+$ denotes the light cone variable. 
The matrix elements in the above equations are defined
as
\begin{eqnarray}
\langle \Gamma (k) \vert~ {\overline O}^{~+}_{(-)} (0,\xi^-,0_\perp) \vert 
\Gamma (k) \rangle \!\!&=&\!\! \sum_{\lambda =0,\pm 1} g^{\lambda \lambda}~ 
\langle \Gamma (k, \epsilon^* (\lambda)) \vert~ O^+_{(-)} (0,\xi^-,0_\perp) 
\vert \Gamma (k, \epsilon(\lambda)) \rangle  ~,
\nonumber \\
\!\! \langle \Gamma(k, \epsilon^*) \vert~ {\delta O}^+_{(-)} (0,\xi^-,0_\perp) 
\vert \Gamma(k, \epsilon) \rangle \!\!&=&\!\! \sum_{\lambda =0, \pm 1} C(\lambda)~
\langle \Gamma(k, \epsilon^* (\lambda)) \vert~ O^+_{(-)} (0,\xi^-,0_\perp) 
\vert \Gamma(k, \epsilon(\lambda)) \rangle ,
\\
\!\!\langle \Gamma(k, \epsilon^*) \vert~ {\Delta O}^+_{(+)5} (0,\xi^-,0_\perp) 
\vert \Gamma (k, \epsilon)  \rangle \!\!&=&\!\! \sum_{\lambda=0, \pm 1} C^\prime (
\lambda) ~\langle \Gamma(k, \epsilon^* (\lambda)) \vert ~ O^+_{(+)5} (0,\xi^-,
0_\perp) \vert \Gamma(k, \epsilon(\lambda)) \rangle  ~. \nonumber
\end{eqnarray}
Using these matrix elements,
the structure functions can be interpreted in terms of the probability
of finding a quark of helicity $h$ in a target photon of helicity $\lambda$
denoted by $f_{{\scriptstyle q}(h)/{\scriptstyle \Gamma}(\lambda)}$ where 
\begin{eqnarray}
f_{\frac {\scriptstyle q(h)} {\scriptstyle \Gamma(\scriptstyle
\epsilon(\lambda))}}(z,\mu^2,\kappa^2) &=& 
\frac{1}{4\pi} \int \d \xi^- e^{-i z ~\xi^- k^+} \langle 
\Gamma(k, \epsilon^*(\lambda)) \vert
\overline \psi (0,\xi^-,0_\perp)~ \gamma^+ \Lambda_h
  ~ \psi (0) \vert \Gamma(k, \epsilon(\lambda)) \rangle ~,
\label{PSF}
\end{eqnarray}
where $\Lambda_h=(1+h~\gamma_5)/2$.  
In terms of $f_{a(h)/\Gamma (\lambda)}$ the structure functions are of the
form
\begin{eqnarray}
F^\Gamma_1 (y) &=& f_{\frac {\scriptstyle a(\uparrow)} {\scriptstyle \Gamma (0)}} - 
f_{\frac {\scriptstyle a(\uparrow)} {\scriptstyle \Gamma (1)}} - 
f_{\frac {\scriptstyle a(\downarrow)} {\scriptstyle \Gamma (1)}}~, \nonumber\\
 b^\Gamma_1(y)&=& 2 \left ( 2 f_{\frac {\scriptstyle a(\uparrow)} {\scriptstyle 
\Gamma (0)}} - f_{\frac {\scriptstyle a(\uparrow)} {\scriptstyle \Gamma (1)}} - 
f_{\frac {\scriptstyle a(\downarrow)} {\scriptstyle \Gamma (1)}} \right )~,
\label{FFD}\\ 
g^\Gamma_1(y) &=& f_{\frac {\scriptstyle a(\uparrow)} {\scriptstyle \Gamma (1)}} - 
f_{\frac {\scriptstyle a(\downarrow)} {\scriptstyle \Gamma (1)}}~. \nonumber
\end{eqnarray}
Interestingly, both $F^\Gamma_{1,2}(y,Q^2,\kappa^2)$ and
$b^\Gamma_{1,2}(y,Q^2,\kappa^2)$ carry nontrivial information about the
photon i.e., parton content of the scalar polarised photon ($
\epsilon_\mu(\lambda=0)$).  This is
absent in the case of real photon as it has only transverse polarisation
states ($\epsilon_\mu(\lambda=\pm 1)$).  Knowing $b^\Gamma_1(y,Q^2,\kappa^2)$
and $F^\Gamma_1(y,Q^2,\kappa^2)$ we can project out the parton content of the
scalar polarised photon.  The authors of Ref.~\cite{HJM} expect that
for $\rho$ meson $b_1^\Gamma \sim {\cal O} (F_1^\Gamma)$ and hence in the
region $0 < \kappa^2 < \Lambda^2$ (where $\Lambda$ is the QCD scale parameter)
the scalar contribution may be substantial.

Now that we know the hadronic structure of the photon in terms of the
helicity states eqn.~(\ref{FFD}), we are in a position to take the
$\kappa^2 \rightarrow 0$ limit and check if we could reproduce the real 
photon results.
The matrix elements (eqn.~(\ref{FFD})) are perturbatively calculable in the
region $\Lambda^2 \ll \kappa^2 \ll Q^2$.
We keep quark 
masses nonzero to exhibit $\kappa^2$ dependence of these operator matrix 
elements.  To regulate ultraviolet divergence we use dimensional 
regularisation and ${\overline{MS}}$ scheme is used to renormalise at the 
scale $\mu_R^2$.  Hence the probability to find a quark with momentum 
fraction $y$ inside transversely polarised ($\epsilon_\mu(\lambda=\pm 1)$) 
and scalar polarised ($\epsilon_\mu(\lambda=0)$) photons is
\begin{eqnarray}
f_{\frac {\scriptstyle q(\uparrow)} {\scriptstyle 
\gamma (\lambda=\pm 1)}} (y,\mu_R^2,\kappa^2) &=& \frac{\alpha} {4 \pi} 
\left [(2 y^2 -2y +1) \ln \frac{{\cal M}_\gamma ^2}{\mu_R^2}- \frac {2 
m^2 y(1-y) }{{\cal M}_\gamma ^2} -\frac {\kappa^2 y(1-y)} {{\cal M}_
\gamma ^2}+ 2 y (1-y) \right ]~, \nonumber \\
f_{\frac {\scriptstyle q(\uparrow)} {\scriptstyle \gamma (\lambda=0)}}
(y,\mu_R^2,\kappa^2) &=& -\frac{\alpha}{2 \pi}  
 \frac{\kappa^2  }{{\cal M}_\gamma ^2} y^2 (1-y)^2 ~,
\end{eqnarray}
where ${\cal M}_\gamma ^2=m^2+\kappa^2y(1-y)$.  It is clear from the above
equation that the regulator (quark mass) going to zero, the quark
content of the scalar polarised photon is $\alpha y(y-1)/(2 \pi)$.
The above exercise proves that the new structure functions
$b^\Gamma_{1,2}(y,Q^2,\kappa^2)$ are of twist two, as the scalar polarised
photon contribution goes as $\kappa^2/{\cal M}^2_\gamma$.  Note that it 
is not a power of $\kappa^2/Q^2$ as in the case of higher twist contributions.  
In the limit $\kappa^2 \rightarrow 0$, $f_{q/\Gamma(\epsilon(0))}$ goes 
to zero and hence we have  $b^\gamma_{1,2}=2 F^\gamma_{1,2}$.  By taking
the limit $\kappa^2 \rightarrow 0$, we would be approaching the nonperturbative 
region and hence the limit is in fact nontrivial.  But since a
real photon ($k^2=0$) is characterised only by $\lambda=\pm 1$, we expect
the scalar contributions should go to zero in the limit $\kappa^2 \rightarrow 
0$.  Now taking the limit $\kappa^2 \rightarrow 0$ in eqns.~(\ref{F1},
\ref{F2}), replacing $\Gamma \rightarrow \gamma$ and equating $b^\gamma_
{1,2} = 2 F^\gamma_{1,2}$, we have
\begin{eqnarray}
F_1^e(x,Q^2) &=& \alpha 
\int \frac{\d \kappa^2} {\kappa^2}~ \int_x^1 \frac{\d y}{y}~
 P_{\gamma e}\left(\frac{x}{y}\right)~ F_1^\gamma(y,Q^2) ~,\\ 
F_2^e(x,Q^2) &=& \alpha 
\int \frac{\d \kappa^2} {\kappa^2}~ \int_x^1 \frac{\d y}{y}~\frac{x}{y}
 ~P_{\gamma e}\left(\frac{x}{y}\right)~ F_2^\gamma(y,Q^2) ~,
\end{eqnarray}
where $P_{\gamma e}$ is the usual Weiz\"acker-William splitting
function for a real photon,
\begin{eqnarray}
P_{\gamma e}\left (\frac{x}{y} \right ) = \frac{y}{x}\left (2-2\frac
{x}{y} + \frac{x^2}{y^2}\right ).
\end{eqnarray}
Thus we have reproduced the real photon results in the limit $\kappa^2 
\rightarrow 0$.

In this section we have used the free field analysis to study the 
physical interpretation of the new singly polarised structure 
function and showed how the virtuality of the target photon alters 
the partonic interpretation of the unpolarised structure function.  
Next we study the QCD corrections to these new structure functions 
and the modified unpolarised structure functions.

\section{QCD corrections: Factorisation method}
>From the free field analysis of the previous section we find that the
hadronic structure of the virtual photon is modified by scalar
polarisation effects.  This modifies the unpolarised structure functions
$F_{1,2}^\Gamma (y,Q^2,\kappa^2)$ while the polarised structure function 
$g_1^\Gamma
(y,Q^2,\kappa^2)$ is unaltered.  Further the virtuality of the photon gives rise
to additional singly polarised structure functions $b^\Gamma_{1-4} (y,Q^2,\kappa^2)$.
In this section we calculate the higher order corrections to the new
twist two structure functions $b_{1,2}^\Gamma (y,Q^2,\kappa^2)$ and the modified
unpolarised structure functions $F_{1,2}^\Gamma (y,Q^2,\kappa^2)$.  As the unpolarised
structure functions $F_{1,2}^\Gamma (y,Q^2,\kappa^2)$ gets modified due to the
virtuality, it is {\it a priori\/} not clear if the Hard Scattering Coefficients
(HSC) are also modified.
For completeness we also
evaluate the higher order correction to $g^\Gamma_1 (y,Q^2,\kappa^2)$.

Higher order corrections (both $em$ and strong) are relevant to the 
study of $\gamma^*~ \Gamma \rightarrow {\rm hadron}$ cross section 
since the photonic corrections go as $\ln Q^2$, whereas the QCD 
corrections turn out to be of leading order, since $f_{q/\Gamma} \sim \ln Q^2
\sim 1/\alpha_s (Q^2)$ (i.e. it compensates for additional power of
$\alpha_s (Q^2)$).  To go beyond the 
leading order we make use of the factorisation approach \cite{CSS} which is a field 
theoretical generalisation of the free 
field analysis.  This approach can be employed for the photon targets 
also since its proof does not depend on the target but depends only on the
underlying theory.  This method ensures
a systematic separation of hard and soft parts, {\it i.e.\/}
\begin{eqnarray}
\!\!\!\!\!\!W_{\mu \nu}^{\gamma^* \Gamma (\epsilon)} (y,Q^2,\kappa^2) 
\!\!&=&\!\! 
\sum_{a,h} \int_y^1 \frac{\d z}{z} f_{{\scriptstyle a(\scriptstyle h)}
/{\scriptstyle \Gamma(\scriptstyle \epsilon)}} (z,\mu_R^2,\kappa^2)~
H^{\mu \nu}_{a(h) \gamma^*} \left (q,zp,\mu_R^2,\alpha_s (\mu_R^2), 
\alpha \right ) + \cdots ~,
\label{FT}
\end{eqnarray}
where the hard part $H$ of the processes are perturbatively calculable
and the summation over $a$ includes partons (quarks and gluons) and free
photons. The `soft' parts are defined below as photon matrix 
elements of bilocal quark, gluon and photon operators,
\begin{eqnarray}
f_{\frac {\scriptstyle q(\uparrow \downarrow)} {\scriptstyle \Gamma(\scriptstyle
\epsilon)}}(z,\mu^2,\kappa^2) &=& 
\frac{1}{4\pi} \int \d \xi^- e^{-i z ~\xi^- k^+} \langle \Gamma (k,\epsilon^*) 
\vert \overline \psi_a (0,\xi^-,0_\perp)~ \gamma^+ \Lambda_\pm
 ~{\cal G}^a_b ~ \psi^b (0) \vert \Gamma (k,\epsilon) \rangle_c ~,
\label{Q}\\
\!\!\! f_{\frac {\scriptstyle {\overline q} (\uparrow \downarrow)} {\scriptstyle 
\Gamma(\scriptstyle \epsilon)}}(z,\mu^2,\kappa^2) &=& \frac{1}{4\pi} 
\int \d \xi^- e^{-i z ~\xi^- k^+} \langle \Gamma(k,\epsilon^*) \vert 
\overline \psi_a (0) ~\gamma^+ \Lambda_\mp ~ {{\cal G}^a_b}^\dagger ~
\psi^b (0,\xi^-,0_\perp) \vert \Gamma(k,\epsilon) \rangle_c ~,
\label{AQ}\\
f_{\frac {\scriptstyle g} {\scriptstyle \Gamma(\scriptstyle \epsilon)}}
(z,\mu^2,\kappa^2) &=& \frac{i}{4\pi z k^+} \int \d \xi^- e^{-i z ~\xi^- k^+} \left 
[\langle k, \epsilon^* \vert F_a^{+\mu} (0,\xi^-,0_\perp)~  {{\cal G}^a_b}
~ F^{+ b}_{\mu} (0) \vert k,\epsilon \rangle_c \right. \nonumber \\
&&\left.  + \langle \Gamma(k, \epsilon^*) \vert F_a^{+\mu} (0)~ {{\cal G}^a_b}^\dagger
~F^{+ b}_{\mu} (0,\xi^-,0_\perp) \vert \Gamma(k,\epsilon) \rangle_c \right ]~,
\label{G}\\
f_{\frac {\Delta \scriptstyle g} {\scriptstyle \Gamma(\scriptstyle \epsilon)}}
(z,\mu^2,\kappa^2) &=& \frac{i}{4\pi z k^+} \int \d \xi^- e^{-i z ~\xi^- k^+} \left 
[ \langle \Gamma(k, \epsilon^*) \vert F_a^{+\mu} (0,\xi^-,0_\perp)~ {\cal G}^a_b ~ 
\widetilde F^{+ b}_{\mu} (0) \vert \Gamma(k,\epsilon) \rangle_c \right . \nonumber \\
&& \left .  - ~\langle \Gamma (k, \epsilon^*) \vert F_a^{+\mu} (0)~{{\cal G}^a_b}^\dagger
~\widetilde F^{+ b}_{\mu} (0,\xi^-,0_\perp) \vert \Gamma(k, \epsilon) \rangle_c \right ] ~,
\label{PG}
\end{eqnarray}
where $\Lambda_\pm = (1 \pm \gamma_5)/2$ and
${\cal G}^a_b = {\cal P}~\exp \left [ig \int_0^{\xi^-} \d\zeta^-
A^+(0, \zeta^-,0_\perp)\right]^a_b$ (${\cal P}$ denotes the path ordering of
gauge fields).  
For $\epsilon(0)$, the $\gamma_5$ term would not be present in the
above definitions.
The above definitions hold for the photon case also wherein the $SU (3)_c$ 
group indices will be absent.  The gauge invariant
definitions of quark, gluon and photon distributions defined above have a
probabilistic interpretation of finding a parton or photon inside the
target photon. 
These matrix elements are in principle calculable if we
treat the photon as a point like perturbative object.
But we know that the photon does not behave like a point like
perturbative object at small energy scales.  Hence we
collect all the higher order effects as well as nonperturbative effects 
inside the matrix elements and treat
them as theoretical inputs.  The fact that they are calculable order
by order in perturbation theory for point like targets such as quarks,
gluons and photons is exploited in the evaluation of HSC.  

The HSCs can be evaluated order by order using
the factorisation formulae by replacing target photon by parton targets 
i.e., quarks, gluons and real photons.  We calculate the HSCs up
to ${\cal O}$$(\alpha^2)$ and ${\cal O}$$(\alpha \alpha_s)$.  Let us first 
concentrate on the quark sector.  The quark sector gets contribution to 
${\cal O}$$(\alpha)$ by $\gamma^* (q)~q(p) \rightarrow q(p^\prime)$
(Fig.~2a), ${\cal O}$$(\alpha^2)$ by $\gamma^* (q)~q (p) \rightarrow 
q(k) ~\gamma(k^\prime)$ and ${\cal O}$ $(\alpha \alpha_s)$ by $\gamma^* 
(q)~q(p) \rightarrow q(k) ~g(k^\prime)$ (Fig.~2b).  For the photonic
corrections, we replace gluon lines by photon lines in Fig.~2b.  From 
the factorisation formulae it is 
clear that the calculation of HSCs involves the
cross sections of the above mentioned processes as well as the matrix
elements given in eqns. (\ref{Q},\ref{AQ}), with target photon replaced 
by  quarks to appropriate orders.  The contributions to various structure 
functions are extracted using the appropriate projection operators 
$P^i_{\mu \nu}$, where $i$ runs over the various unpolarised, singly 
polarised and polarised structure functions.  These projection operators 
are given in the Ref.~\cite{PR,PM}.  We define $W_{\gamma^* q}^i = 
P^i_{\mu \nu} W^{\mu \nu}_{\gamma^* q}$ (i.e., in eqn.~(\ref{FT})
replace the target photon $\Gamma$ with quark and evaluate the evaluate
to ${\cal O} (\alpha \alpha_s)$).  This corresponds to evaluation of  
the bremstrahlung diagrams given in Fig.~2b. At large $Q^2$, we get
\begin{eqnarray}
W^{F_1}_{\gamma^* q}(z,Q^2) &=& 2  f_c~ \alpha ~\left \{ -\left (\frac{1+z^2}
{1-z} \right )_+ \ln \beta_g - 2 ~\frac{1+z^2}{1-z} ~\ln z + (1+z^2) \left (
\frac{\ln (1-z)}{1-z} \right)_+ \right .\nonumber \\
&& \left . + 2z + 1 -\frac{3}{2} \frac{1}{(1-z)_+} -
\delta (1-z) \left( \frac{9}{4} + \frac{2 \pi^2}{3} \right ) \right \}
~,
\nonumber\\
W_{\gamma^* q}^{F_2}(z,Q^2) &=&  2z \left \{ W_{\gamma^* q}^{F_1} (z,Q^2) + 2 \alpha ~
f_c ~ 2z \right \}~,
\nonumber\\
W_{\gamma^* q}^{b_1} (z,Q^2)&=& 2 ~W_{\gamma^* q}^{F_1} (z,Q^2)  ~,\label{CSB1}\\
W_{\gamma^* q}^{b_2}(z,Q^2) &=& 2 ~ W_{\gamma^* q}^{F_2}(z,Q^2)  ~, \nonumber\\
W_{\gamma^* q}^{g_1}(z,Q^2)&=&  W_{\gamma^* q}^{F_1} (z,Q^2) - 2 \alpha ~f_c  ~ (1-z) ~,
\nonumber
\end{eqnarray}
where $z=Q^2/2p \cdot q$~, $\beta_g=m_g^2/Q^2$, $f_c =(4 \alpha_s/3, \alpha)$ is the coupling
factor depending on gluon or photon bremstrahlung respectively and the
subscript + denotes the `+ function' regularisation of the singularity as
$z \rightarrow 1$.  For the photons $m_g$ will be replaced by
$m_\gamma$. 
To avoid the mass singularity, we have kept the gauge bosons massive.  If
quark masses are also kept nonzero, then there would be both logarithmic and power 
singularities as these masses go to zero simultaneously.  If one of the
prescriptions is chosen, say massive gauge boson prescription, the singularities
boil down to logarithmic singularities with an additional constant part which
depends on the prescription.  In addition there are some mass singularities
coming from the virtual diagrams which are exactly canceled by  
those which arise from
regulating the bremstrahlung diagram in the limit $z \rightarrow 1$ (`+ 
 function').  We have considered the massive gauge boson prescription
which has not so far been considered in the literature while calculating 
the corrections to structure functions.  This is a convenient
choice since we are dealing with the virtual photons.  It has been customary 
to consider the massive quark prescription or the dimensional regularisation 
method to deal with the infrared (IR) singularities.

For the structure functions $F^\Gamma_{1,2}(y,Q^2,\kappa^2)$ and 
$b^\Gamma _{1,2}(y,Q^2,\kappa^2)$ the relevant matrix elements 
are obtained by replacing $\Gamma$ in eqns.~(\ref{Q},\ref{AQ}) by 
quarks, where only the vector operator will contribute.  In the case of 
polarised structure function $g^\Gamma_1(y,Q^2,\kappa)$ 
the axial vector operator will contribute (eqns.~(\ref{Q},
\ref{AQ})).  The contributing matrix elements are shown in Fig.~3.  
The Feynman rules for the eikonal lines and vertices are given in 
Ref.~\cite{CSS}.  We regulate the ultraviolet (UV) divergences 
appearing in these diagrams using dimensional regularisation and 
keep gauge boson masses nonzero to regulate the mass singularities.  
Here too there is a similar cancellation of mass singularities among 
the virtual and real diagrams, leaving a logarithmic singularity and 
a prescription dependent constant as given below
\begin{eqnarray}
f^{(1)}_{\frac {\scriptstyle q} {\scriptstyle q}} (z,\mu_R^2)\!&=&\!  
\frac{f_c}
{4\pi}\left \{ \left ( \frac{1+z^2}{1-z} \right )_+ \ln \beta^\prime + 
\frac{1+z^2}{1-z} \ln z + 2(1-z)- \delta(1-z) \left ( \frac {9}{4} - 
\frac{\pi^2}{3} \right ) \right \} ,\nonumber \\
f^{(1)}_{\frac {\Delta \scriptstyle q} {\scriptstyle q(\scriptstyle h)}}
(z,\mu_R^2) \!\!\! &=& \!\!\! h~ f^{(1)}_{\frac {\scriptstyle q} 
{\scriptstyle q}}(z,\mu_R^2) ~,
\end{eqnarray} 
where $\beta_g^\prime = {m_g^2}/{\mu^2_R}$ and $\mu_R$ is the renormalisation
scale. The superscript $(1)$ in the above equations denotes that they are 
evaluated to order $\alpha$ or $\alpha_s$ as the case may be.
This equivalence among the polarised and unpolarised matrix elements 
does not hold if we keep the quark masses also nonzero.
Substituting the above matrix elements and cross sections (eqn.~(\ref{CSB1}))
in the factorisation formulae for quark sector, we obtain
\begin{eqnarray}
H^{F_1}_{q \gamma^*}(z,Q^2) &=&  2 \alpha ~f_c~\left \{ \left (\frac{1+z^2}
{1-z} \right )_+ \ln \frac{Q^2}{\mu_R^2} -~\frac{1+z^2}{1-z} ~\ln z + 
(1+z^2) \left ( \frac{\ln (1-z)}{1-z} \right)_+ \right .\nonumber \\
&& \left . + 3 -\frac{3}{2} \frac{1}{(1-z)_+} -
\delta (1-z) \left( \frac{9}{2} + \frac{\pi^2}{3} \right ) \right \} ~,\nonumber \\
H^{F_2}_{q \gamma^*}(z,Q^2) &=& 2z \left[H^{F_1}_{q \gamma^*}(z,Q^2) +  2 f_c ~\alpha~
2z \right] ~,\nonumber \\
H^{b_1}_{q \gamma^*}(z,Q^2) & =& 2 H^{F_1}_{q \gamma^*}(z,Q^2) ~,\\
H^{b_2}_{q \gamma^*}(z,Q^2) & =& 2 H^{F_2}_{q \gamma^*}(z,Q^2) ~,\nonumber \\
H^{g_1}_{q \gamma^*}(z,Q^2) & =& H^{F_1}_{q \gamma^*}(z,Q^2) +  2 \alpha ~f_c
(z-1)~\nonumber .
\end{eqnarray} 
Note that the mass term in the logarithmic and the prescription dependent 
constant term cancel among the cross section $W^i_{\gamma^* q}~ (z,Q^2)$ and 
the matrix 
element $f_{\scriptstyle a/\scriptstyle b}~ (z,Q^2)$, leaving behind the HSC 
independent of gauge boson mass.

Next we will discuss the corrections coming from the gluon initiated
subprocesses.  The contributing subprocesses are 
$\gamma ^*(q)~ g(p) \rightarrow q(k)~ \bar q(k^\prime)$ (Fig.~4) and  
$\gamma^* (q) ~\gamma (p) \rightarrow q(k) ~\bar q(k^\prime)$.  To calculate 
the gluonic and photonic HSCs we need in addition to the above cross
sections, the matrix element (eqns.~(\ref
{G},\ref{PG})) between gluon states and between photon states respectively. 
The contributions to various structure 
functions from the cross section, at large $Q^2$ are
\begin{eqnarray}
W_{g \gamma^*}^{F_1} (z,Q^2) &=& 8~\alpha~ f_c \left \{ (2z^2-2z+1)
\left (\ln \frac{{\cal M}_g^2}{Q^2} + \ln \frac{z}{1-z} \right ) 
\right . \nonumber \\
&& \left . - \frac{p^2} {{\cal M}_g^2}z(1-z) 
- \frac{2m^2 }{{\cal M}_g^2}z(1-z) + 2z^2 -2z +1 \right \} ~, 
\nonumber\\
W_{g \gamma^*}^{F_2} (z,Q^2) &=& 2z \left \{ W_{g \gamma^*}^{F_1} (z,Q^2) +
8~ \alpha~f_c~(4z^2 - 4z)  \right \} ~,
\nonumber \\ 
W_{g \gamma^*}^{b_1} (z,Q^2) &=&  2~ \left (W_{g \gamma^*}^{F_1} (z,Q^2) 
- 16 ~\alpha~f_c \frac{p^2}{{\cal M} _g^2} z^2 (1-z)^2 \right)~,\\
W_{g \gamma^*}^{b_2} (z,Q^2) &=& 2~ \left (W_{g \gamma^*}^{F_2} (z,Q^2) 
- 32 z~\alpha~f_c \frac{p^2}{{\cal M} _g^2} z^2 (1-z)^2 \right )~,
\nonumber \\
W_{g \gamma^*}^{g_1} (z,Q^2) &=& 4~ \alpha~ f_c  \left \{ (1-2z)
\left (\ln \frac{{\cal M}_g^2}{Q^2} + \ln \frac{z}{1-z} 
- \frac{p^2} {{\cal M}_g^2} z(1-z) + 1 \right )
+ \frac {2m^2 }{{\cal M}_g^2} (1-z) \right \} ~,
\nonumber
\end{eqnarray} 
where ${\cal M}_g^2 \equiv m^2 - p^2 z(1-z)$.
Observe that the new singly polarised structure functions $b^\Gamma_
{1,2} (z,Q^2,\kappa^2)$ differs from $F^\Gamma_{1,2}(z,Q^2,\kappa^2)$ by 
a term  $(2p^2 z^2(1-z)^2/ {\cal M}_g^2)$.  This extra term 
is due to the gauge boson of
scalar polarisation, as a result of the off-shell nature of the gauge boson.
In the parton model, care should be taken in using the above result for the 
subprocess cross sections because the unphysical scalar polarisation gluon
should not be considered.  In this procedure to calculate the HSCs, this
step is just a technique and we have no reason to avoid the scalar polarisation.
As we go along, it will become clear that this unphysical degree of freedom 
will have no effect on the HSC.  The relevant matrix elements are
\begin{eqnarray}
\!\!\!\!\!f^{(1)}_{\frac {\scriptstyle q} {\scriptstyle g}}(z,\mu_R^2)\!\!\! &=&\!\!\!
\frac{f_c}
{4 \pi} \left [(2 z^2 -2z +1) \ln \frac{{\cal M}_g^2}{\mu_R^2}- \frac
{2 m^2 }{{\cal M}_g^2} z(1-z) -\frac {p^2} {{\cal M}_g^2} z(1-z)+ 2 z 
(1-z) \right ]~, \nonumber \\
f^{(1)}_{\frac {\delta \scriptstyle q} {\scriptstyle g}}(z,\mu_R^2) &=& 
f^{(1)}_{\frac {\scriptstyle q} {\scriptstyle g}}(z,\mu_R^2)
- \frac{f_c}{4 \pi} \left [2 \frac{p^2}{{\cal M}_g^2} z^2 (1-z)^2 \right ]~,\\
f^{(1)}_{\frac {\Delta \scriptstyle q} {\scriptstyle g (h)}}(z,\mu_R^2)\!\! &=&\!\! 
h~\frac{f_c}{4 \pi} \left [(1 -2z) \ln \frac{{\cal M}_g^2}
{\mu_R^2} + \frac{p^2}{{\cal M}_g^2} z(1-z) \right ] ~,
\nonumber  
\end{eqnarray}
which are evaluated from the cut diagrams shown in Fig.~5.  The details of the
calculation can be found in Ref. \cite{PR} for polarised case and
in Ref. \cite{PM} for the unpolarised case.  Substituting the cross section and matrix elements in the 
factorisation formulae, we get
\begin{eqnarray}
H_{g \gamma^*}^{F_1}(z,Q^2) &=& 8 \alpha f_c \left [ (2z^2 -2z
+1) \left ( - \ln \frac{Q^2}{\mu_R^2} + \ln \frac{z}{1-z} \right ) + 4z^2 -4z +
1 \right ] ~,
\nonumber\\ 
H_{g \gamma^*}^{F_2}(z,Q^2) &=& 2z \left [ H_{g \gamma^*}^{F_1} (z,Q^2) + 8
\alpha f_c(4z^2 -4z) \right ] ~,
\nonumber\\
H_{g \gamma^*}^{b_1}(z,Q^2) &=& 2 H_{g \gamma^*}^{F_1}(z,Q^2) ~,\\
H_{g \gamma^*}^{b_2}(z,Q^2) &=& 2 H_{g \gamma^*}^{F_2}(z,Q^2) ~,
\nonumber\\
H_{g \gamma^*}^{\scriptstyle g_1}(z,Q^2) &=& 4\alpha f_c\left[ (2z-1) \left (\ln 
\frac{Q^2} {\mu_R^2} - \ln\frac{z}{1-z} \right ) -4z+3\right] ~.
\nonumber
\end{eqnarray}
As expected the mass terms cancel among cross section and 
matrix element, so also the scalar gauge boson contribution.  
Substituting the calculated HSCs in eqn.~(\ref{FT}) one 
gets the QCD corrected $W^\Gamma_{\mu \nu} (y,Q^2,\kappa^2)$ for 
finite virtual target photon mass.  Choose the renormalisation 
scale $\mu_R^2 = Q^2$ so that the $Q^2$ dependency of the HSCs 
will be transferred to the strong coupling constant and the 
parton distribution function $f_{a/\Gamma} (z,\mu_R^2=Q^2,\kappa^2)$.  
At every order in $\alpha_s (Q^2)$, the $\ln Q^2$ growth of 
$f_{a/\Gamma} (Q^2)$ is compensated for by the coupling constant.
The photonic operator also contributes in the same order.  But 
due to the $\ln Q^2$ growth of $f_{q/\Gamma} (Q^2)$, the photon
structure tensor $W^\Gamma_{\mu \nu}$ grows as $\ln Q^2$ to
leading order $(\alpha_s^0)$. The first moments of the gluonic 
and photonic HSC of $g_1^\Gamma (z,Q^2,\kappa^2)$ vanish
and hence are not corrected by these operators.  So the first 
moment of $g_1^\Gamma (z,Q^2,\kappa^2)$
is proportional to only quark field operators i.e. $f_{\Delta q/
\Gamma} (Q^2,\kappa^2)$.  
For off-shell 
target photons this quantity is nonzero.

\section{Conclusion}
We have analysed the $\gamma^* \Gamma \rightarrow X$ subprocess 
of the $e^+ e^- \rightarrow e^+ e^- X$ process in the DIS limit. 
Our analysis correctly includes the virtuality of the target photon
by taking into account the contribution coming from the scalar
polarisation.
The virtuality of the target photon gives rise to four new singly
polarised structure functions.  Two of these new structure functions are
found to be of twist two.  Hence we find that the unpolarised 
cross section is modified by these twist two structure functions to 
leading order.  In addition the usual $e \rightarrow \gamma$ splitting functions are altered
by the scalar polarisation of the virtual target photon.  

Using the
free field analysis we have studied the physical interpretation of
the new structure functions $b_{1,2}^\Gamma (y,Q^2,\kappa^2)$ and
we also find that the physical interpretation of unpolarised
structure functions $F_{1,2}^\Gamma (y,Q^2,\kappa^2)$ are modified
by the scalar polarisation.  In the process we have obtained relations among
various structure functions.  The free field analysis is useful 
to show that our results are consistent with the existing real 
photon results in the limit $\kappa^2 \rightarrow 0$.  

In addition we have also systematically
computed various QCD and QED contributions to these structure
functions using the factorisation method. We have redone this because 
the old results are modified by the
existence of extra polarisation state $viz.$ parton distribution function 
in a scalar polarised photon.  The differential cross
section is found to grow as $\ln Q^2$ while higher order QCD
corrections are found to contribute to leading order.  Interestingly,
the first moment of $g^e_1 (y,Q^2)$ is proportional only to
the first moment of the quark operator $f_{\Delta q/\Gamma} (Q^2,\kappa^2)$.
\vspace{.5cm}

\noindent
{\Large Acknowledgements}

We thank G T Bodwin for clarifying some points regarding the 
regularisation scheme adopted in the context of factorisation
method.  Thanks are due to R M Godbole, H S Mani, M V N Murthy, 
J Pasupathy and R Ramachandran for useful discussion. One of us
PM would like to thank M Gl\"uck and E Reya for useful discussion. 
We acknowledge the use of symbolic manipulation packages such as 
FORM and MACSYMA.
\vspace{.5cm}

\eject
\noindent
{\Large Appendix}

The unpolarised cross section in eqn.~(\ref{UPCS}) can be derived from 
eqn.~(\ref{TCS}) by summing over the polarisation states of the positron, as
\begin{eqnarray}
\frac{d \sigma^{s_1 (\uparrow +\downarrow)}}{d x ~ d Q^2} = \frac{\alpha^2}{4 x^2 s^2 Q^2}~ 
L^{\mu \mu^\prime} (q,p_1,s_1)~ \left (W^e_{\mu \mu^ \prime} (q,p_2,\uparrow)
+ W^e_{\mu \mu^ \prime} (q,p_2,\downarrow) \right) ~.
\end{eqnarray}
>From eqn.~(\ref{EST}), we have
\begin{eqnarray}
W^e_{\mu \mu^\prime}(x,Q^2,\uparrow) + W^e_{\mu \mu^\prime}(x,Q^2,\downarrow)
&=&\alpha x \int \frac{{d \kappa^2}}{\kappa^2}~ \int^1_x \frac{d y}{y^2} 
\left (L^{\nu \nu^\prime} (k,p_2,\uparrow) + L^{\nu \nu^\prime} (k,p_2,
\downarrow) \right )\nonumber \\ 
&&\times \sum_{\lambda=0, \pm 1} g^{\lambda \lambda}~ \epsilon^*_{\nu} (k,\lambda)~
\epsilon _{\nu^\prime} (k,\lambda)~ W^\Gamma_ {\mu \mu^\prime} (q,k,\lambda)~.
\label{SPT}
\end{eqnarray}
Since the combination 
$(L^{\nu \nu^\prime} (k,p_2,\uparrow) + L^{\nu \nu^\prime} (k,p_2,\downarrow))$ 
is symmetric in $\nu \nu^\prime$ only the symmetric part of 
$\epsilon_\nu^* (k,\lambda) \epsilon_{\nu^\prime} (k,\lambda)$ gets projected.  Hence 
\begin{eqnarray}
\sum_{\lambda=0 \pm 1} g^{\lambda \lambda} \epsilon^*_\nu (k,\lambda) 
\epsilon_\nu^\prime (k,\lambda) ~W^\Gamma_{\mu \mu^\prime} (\lambda) 
&=& \epsilon^*_\nu (k,0) \epsilon_{\nu^\prime} (k,0) ~W^\Gamma_{\mu \mu^
\prime} (0) \nonumber \\
&& - \epsilon^*_\nu (k,1) \epsilon_{\nu^\prime} (k,1) 
\left (W^\Gamma_{\mu \mu^\prime} (1) + W^\Gamma_{\mu \mu^\prime} (-1)
\right ), 
\end{eqnarray}
where we have used the fact that the symmetric part of $\epsilon^* (k,1) 
\epsilon (k,1) =\epsilon_\nu^* (k,-1) \epsilon_{\nu^\prime} (k,-1)$.   In 
the above equation $W^\Gamma_{\mu \mu^\prime} (0)$ and $W^\Gamma_{\mu \mu
^\prime} (1) + W^\Gamma_{\mu \mu^\prime} (-1)$ are both symmetric and 
hence consists of two parts $viz.$ unpolarised and singly polarised (eqn.
~(\ref{PST})).  These are denoted by the superscripts $UP$ and $SP$. 

Now we make use of the following properties of the unpolarised and
singly polarised sector (eqn.~(\ref{PST})),--- (i)the unpolarised 
part is independent of the polarisation of the photon
\begin{eqnarray}
W^{\Gamma (UP)}_{\mu \mu^\prime} (0) =
W^{\Gamma (UP)}_{\mu \mu^\prime} (1) =
W^{\Gamma (UP)}_{\mu \mu^\prime} (-1), 
\end{eqnarray}
and (ii)sum of polarisation of the singly polarised part is
\begin{eqnarray}
\sum_{\lambda=0 \pm 1} g^{\lambda \lambda} W^{\Gamma (SP)}_
{\mu \mu^\prime} (\lambda) =0 . 
\end{eqnarray}
The above two properties implies 
\begin{eqnarray}
\sum_{\lambda=0 \pm 1} g^{\lambda \lambda} W^{\Gamma }_
{\mu \mu^\prime} (\lambda) = -W^{\Gamma (UP)}_{\mu \mu^\prime}, 
\end{eqnarray}
and 
\begin{eqnarray}
\sum_{\lambda=0 \pm 1} C(\lambda) W^\Gamma_{\mu \mu^\prime} (\lambda) 
= W^{\Gamma (SP)}_{\mu \mu^\prime}(0). 
\end{eqnarray}
Making use of these properties we get
\begin{eqnarray}
\sum_{\lambda=0\pm 1} g^{\lambda \lambda}\epsilon^*_\nu (k,\lambda) 
\epsilon_\nu^\prime (k,\lambda) W^\Gamma_{\mu \mu^\prime} (\lambda) 
&=& -\sum_{\lambda=0, \pm 1} g^{\lambda \lambda} \epsilon^*_\nu (k,\lambda) 
\epsilon_\nu^\prime (k,\lambda) \sum_{\lambda=0, \pm 1} g^{\lambda \lambda} 
W^\Gamma_{\mu \mu^\prime} (\lambda) \nonumber \\
&& + \frac{1}{2} \sum_{\lambda=0 \pm 1}
C(\lambda) \epsilon^*_\nu (k,\lambda) \epsilon_\nu^\prime (k,\lambda) 
\sum_{\lambda=0, \pm 1} C(\lambda) W^\Gamma_{\mu \mu^\prime} (\lambda). 
\end{eqnarray}
Substituting the above equation in eqn.~(\ref{SPT}) we reproduce the 
unpolarised cross section eqn.~(\ref{UPCS}).

As the scalar polarisation of the virtual photon does not affect the 
polarised cross section, the derivation of the polarised cross section 
is the same as in the real photon case and hence is not repeated.


\eject

\eject
{\Large \bf Table.~1}
\begin{list}
{}{\setlength{\labelwidth}{20mm}}
\item Table.~1 
Various structure functions related to scaling functions and their $n^{\rm
th}$ moments.
\end{list}
\vspace{1cm}

\begin{tabular}{|c||c|c|} \hline
Structure Function & Scaling Functions & Moments \\ \hline \hline 
$F_1(y)$ & $\widetilde A (y)$ & $\frac{(-i)^{n-1}}{4} A_{n-1}$ \\ \hline 
$F_2(y)$ & $2y \widetilde A (y)$ & $\frac{(-i)^n}{2} A_n$ \\ \hline \hline
$b_1(y)$ & $- \frac{i}{2} y \frac{\partial^2} {\partial y^2} \widetilde 
B (y)$ & $\frac{(-i)^{n-1}}{4} B_{n-1}$ \\ \hline 
$b_2(y)$ & $- i y^2 \frac{\partial^2} {\partial y^2} \widetilde 
B (y)$ & $\frac{(-i)^n}{2} B_n$ \\ \hline 
$b_3(y)$ & $ - i y \frac{\partial} {\partial y} \widetilde B (y)$ & 
$-\frac{(-i)^n}{2(n+1)} B_n$ \\ \hline 
$b_4(y)$ & $- i \widetilde B (y)$ & $\frac{(-i)^n}{2n(n+1)} B_n$ \\ \hline
\hline 
$g_1(y)$ & $- \frac{1}{2} y \frac{\partial} {\partial y} \widetilde 
C (y)$ & $\frac{(-i)^{n-1}}{4} C_{n-1}$ \\ \hline 
$g_2(y)$ & $\frac{1}{2} \widetilde C (y) + \frac{1}{2} y \frac{\partial} 
{\partial y} \widetilde C (y)$ & $- \frac{(-i)^{n-1}}{4} \frac{n-1}{n} 
C_{n-1}$ \\ \hline 
\end{tabular}\vspace{.5cm}\\
\eject

{\Large \bf Figure Captions}
\begin{list}
{}{\setlength{\labelwidth}{20mm}}
\item  [Fig 1.]
The process $e^+ e^- \rightarrow e^+ e^- X$ via the
photon-photon interaction.

\item  [Fig 2.]
(a) Born diagram, (b) Next to leading order corrections to it.

\item  [Fig 3.]
The contribution to the matrix element $f_{{\scriptstyle q}/
{\scriptstyle q}}$ and $f_{{\Delta \scriptstyle q}/{\scriptstyle q}}$ 
up to ${\cal O} (\alpha_s)$.

\item  [Fig 4.]
The gluon (photon) produced in the target photon 
interacting with the probe photon at a higher order.

\item  [Fig 5.]
The ${\cal O} (\alpha_s)$ contribution to the matrix
element $f_{{\scriptstyle q}/{\scriptstyle g}}$ and $f_{{\Delta 
\scriptstyle q}/{\scriptstyle g}}$. The vertex is $\gamma^+$ for 
unpolarised and $\gamma^+ \gamma_5$ for polarised.  Here only the 
nonvanishing diagrams are given.

\end{list}

\end{document}